# Hierarchy and assortativity as new tools for affinity investigation: the case of the TBA aptamer-ligand complex

Rosella Cataldo[1,3*], Eleonora Alfinito[2*], Lino Reggiani[1]

[1]*Dipartimento di Matematica e Fisica "Ennio de Giorgi", Università del Salento, via Monteroni, Lecce, Italy*
[2]*Dipartimento di Ingegneria dell`Innovazione. Università del Salento, via Monteroni, Lecce, Italy*
[3]*Istituto Nazionale di Fisica Nucleare, INFN, Sezione di Lecce, Italy*

**Abstract**
Aptamers are single stranded DNA, RNA or peptide sequences having the ability to bind a variety of specific targets (proteins, molecules as well as ions). Therefore, aptamer production and selection for therapeutic and diagnostic applications is very challenging. Usually they are *in vitro* generated, but, recently, computational approaches have been developed for the *in silico* selection, with a higher affinity for the specific target. Anyway, the mechanism of aptamer-ligand formation is not completely clear, and not obvious to predict. This paper aims to develop a computational model able to describe aptamer-ligand affinity performance by using the topological structure of the corresponding graphs, assessed by means of numerical tools such as the conventional degree distribution, but also the rank-degree distribution (hierarchy) and the node assortativity. Calculations are applied to the thrombin binding aptamer (TBA), and the TBA-thrombin complex, produced in the presence of $Na^+$ or $K^+$. The topological analysis reveals different affinity performances between the macromolecules in the presence of the two cations, as expected by previous investigations in literature. These results nominate the graph topological analysis as a novel theoretical tool for testing affinity. Otherwise, starting from the graphs, an electrical network can be obtained by using the specific electrical properties of amino acids and nucleobases. Therefore, a further analysis concerns with the electrical response, which reveals that the resistance sensitively depends on the presence of sodium or potassium thus posing resistance as a crucial physical parameter for testing affinity.

## 1. Introduction

Medicine is addressing more and more efforts toward prevention, personalization of cure and the search of less invasive and point-of-care diagnostic methods. Accordingly, the development of new analysis techniques and therapies is a challenging task for researches belonging to all the branches of science, from biology to informatics and from physics and chemistry to engineering [1-4].

Since their discovery in 1990, aptamers have been extensively used in the development of various bioanalytical assays, giving astonishing results. They are small fragments of single stranded DNA or RNA, artificially produced with high binding affinity and specificity, to perfectly adapt to an assigned ligand (from small molecules to large proteins). The technique used for their selection, also known as SELEX (Systematic Evolution of Ligands by EXponential Enrichment) [5] seems so powerful to produce, in principle, an aptamer for each specific pathogen or macromolecule found to be at the origin of a disease. SELEX is *in vitro* technique based on three steps: incubation, evolution and amplification. The whole process is repeated for various rounds until the random library is enriched with the sequences of higher affinity for the target [2]. On the other hand, computational (*in silico*) less expensive approaches have been widely developed for aptamer identification and optimization [6]. In general, they 'dock' small molecules into the structures of macromolecular targets and 'score' their potential complementarity to binding sites. In the last 20 years, nearly 5000 publications focused on aptamers for analytical developments, showing interesting results in selecting efficient aptamers, by using both biochemical and computational techniques [1, 2, 4, 7].

Anyway, despite of all these efforts, many problems concerning the mechanism of aptamer-ligand binding remain unsolved.

This paper proposes a theoretical/computational approach, within the emergent science of Proteotronics [3] as a novel tool to reproduce chemical affinity in aptamer-ligand complexes. It is able to interpret the electrical responses observed in experiments [8–13], and to be predictive of novel results. The modelling, initially proposed to describe the electrical responses and topological features of proteins (sequences of amino acids) when inserted in electronic devices, has been recently applied to the thrombin-binding aptamer (TBA) (sequence of nucleobases) [14]. By using this technique, some relevant results, obtained with X-ray spectroscopy [15] and electrochemical impedance spectroscopy (EIS) measurements [13] have been well reproduced.

*Corresponding Authors:
eleonora.alfinito@unisalento.it
rosella.cataldo@unisalento.it

The aptamer, alone and complexed with the protein, is represented by a complex network. Since the mid 1990s, network science was engaged in the effort to characterize network structure and function, in which complexity arises as an emerging property of the macroscopic behaviour of a system of interacting elements [16–18]. Recently, network architectures have been applied in several empirical studies, such as structural and functional human and other animals brain networks [19], as well as the analysis of regional economic and social systems of innovation [20], showing that network organization favours the production and diffusion of knowledge [21].

The statistical properties of complex networks are usually investigated by using several tools, able to measure both global and local structures. In the present paper, we focus on: (i) the degree distribution, describing the probability that a randomly chosen vertex has degree $k$, (ii) the degree–rank distribution, giving the relationship between the degree and the rank of the degree sequence and, (iii) the so-called assortativity [21–24]. The structure of relations is assortative when highly (poorly) connected nodes tend to be linked to other high (poor) degree nodes, and disassortative when highly (poorly) connected nodes tend to be linked to other poor (high) degree nodes. Therefore, the level of network assortativity gives a formal representation of the way information (affinity performance) flows between central and more peripheral nodes. Discussing the results obtained by means of those indicators, valuable information about affinity performances in an aptamer-ligand complex can be derived and validated by comparison with experimental findings.

The development of the model encompasses two fundamental steps: I. the building of a graph analogue; II. the building of an interaction network. Both steps preserve the macromolecule structure and therefore the interaction features (here electrical), reflect the fundamental *structure & function* paradigm in biology. Finally, the model is able to resolve structures with full-blown different affinities.

The model is applied to the small 15-mer TBA, whose ability in the inhibition of the enzyme thrombin, involved in several blood diseases, is well known. Due to the large literature dealing with thrombin and its aptamers, it is the most commonly used system to demonstrate the proof-of-concept of the aptamer-based affinity assays.

In particular, we are in a position to foresee the reduced affinity of the TBA-thrombin complex, when produced in a solution containing $Na^+$, with respect to the same compound, produced in a solution containing $K^+$. Furthermore, the resistance variation observed in EIS measurement can be also well reproduced [14].

## 2. MATERIALS AND METHODS

### 2.1 Topological description

For the aptamer 5'-GGT TGG TGT GGT TGG-3` (TBA), we consider the Protein Data Bank-PDB-entries [25] related to:

*a*. the aptamer in its native state, i.e. its lowest free energy state;

*b*. the aptamer in its active form, i.e. the aptamer with the structure deformed due to the binding but deprived of the protein, in the presence of both $K^+$ and $Na^+$;

*c*. the aptamer-enzyme complex, in the presence of both $K^+$ and $Na^+$.

PDB format provides a standard representation for macromolecular structure data, derived from X-ray diffraction and NMR studies. It contains at least one model of the so-called tertiary structure (the 3D arrangement of the macromolecule, i.e. the coordinates of all the atoms). In the present analysis we select the carbon atoms $C_1$ or $C_\alpha$ to represent the position of each nucleobase or amino acid. Therefore, hereafter, these carbon atoms will be identified with the name *elementary bricks* (*b*).

Structure *a*. is an NMR product, available at the entry 148D [26]. Structures *b-c* are available at the entries: 4DII and 4DIH and correspond to the X-ray structures of the complex between human alpha thrombin and TBA in the presence of $K^+$ and $Na^+$ ions, respectively [15,25]. Those last structures are particularly important in evaluating affinities, since it is well known that ions play an important role in stabilizing the 3D structure of TBA, the so-called G-quadruplex. In particular, by adding $K^+$ ions, the resulting G-quadruplex is more stable and an increased inhibitory activity on thrombin is found with respect to compounds obtained in the presence of $Na^+$ ions [15].

*Corresponding Authors:
eleonora.alfinito@unisalento.it
rosella.cataldo@unisalento.it

The numerical procedure starts by building the networks (graphs) corresponding to the 3D structures of the aptamer, alone and/or complexed. Extensive description of graph analysis can be found in the literature [16 – 18]. We represent a network as a graph $G(N,L)$, with node set $N$ and link set $L$.

The input data are:
i. The backbone 3D structures (*a-c*), i.e. the coordinates of each $b_i$ ($1 \leq i \leq N$) elementary brick;
ii. The value of the cut-off radius $R_C$.

The cut-off radius $R_C$ is a free parameter, whose tuning produces a graph more connected at increasing values [3].

Accordingly, starting from the backbone, all the Euclidean distances between the couples of $b_i$, $b_j$ elementary bricks are calculated, creating a distance matrix. From the distance matrix, the graph description of the network is represented through its adjacency matrix $A$ of size $N \times N$, with elements

$a_{i,j}$ = 1 (there is a link), if the distance between node $i$ and $j$ is less than the assigned cut-off radius $R_C$

= 0 (there is no link), if the distance between node $i$ and $j$ is greater than the assigned cut-off radius $R_C$.

We assume that no self-loops exist (hence $a_{i,i} = 0$) and no overlapping links, i.e. there cannot be more than one link between $i$ and $j$ (simple graph). Furthermore, graph is undirected, i.e. $a_{i,j} = a_{j,i}$, i.e., $A = A^T$.

Figure 1 shows the network topology of the considered structures with $R_C = 11.3$ Å; this value corresponds to the max resolution in resistance, see next paragraph 2.2. In Table 1, the topological characteristics of the considered networks are presented; those structures encompass about 3500 links but, for a very large cut-off value, i.e. $R_C = 30$ Å, they present about 25000 links.

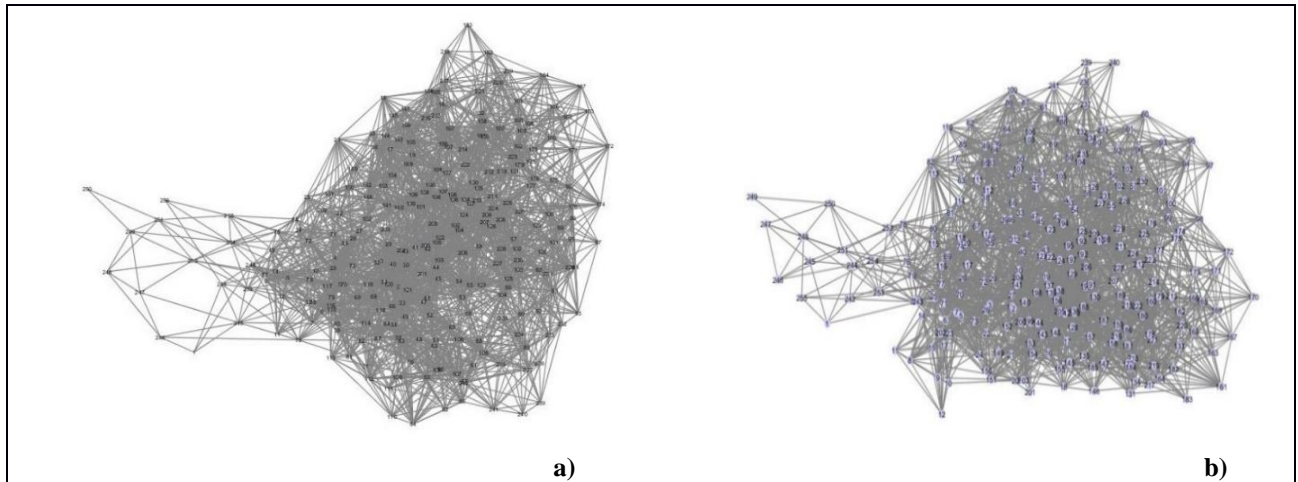

a) b)

**Figure 1.** *Network topology of the considered structures: a) 4DII and b) 4DIH, with $R_C = 11.3$ Å.*

|      | # of nodes | # of links | MD    | C     | C/Cr | L    | L/Lr |
|------|------------|------------|-------|-------|------|------|------|
| **4DII** | 256        | 3296       | 25.75 | 0.568 | 5.6  | 2.82 | 1.43 |
| **4DIH** | 255        | 3365       | 26.39 | 0.569 | 5.5  | 2.78 | 1.42 |

**Table 1**. *Average path length (L) and the clusterization coefficient C for the same networks in Figure 1, together with the same values normalized ($C_r$ and $L_r$) for a random lattice with equal number of nodes N and equal mean degree (MD).*

*Corresponding Authors:
eleonora.alfinito@unisalento.it
rosella.cataldo@unisalento.it

In Table 1, the average path length L and the clusterization coefficient C are calculated, for the same networks in Figure 1, together with the same values normalized to $C_r$ and $L_r$ (which are the clusterization coefficient and the average path length calculated for a random lattice with equal number of nodes N and equal mean degree, MD) [27].

The result was a large C (far greater than $O(N^{-1})$ which would be typical of random graphs and similar to $C_r$) and a relatively low average path length, which is compatible with the hypothesis of a small-world network [18], with significant clusterization. The small world network approach is attractive from the conceptual perspective, providing better understanding of the stability of protein topology, both in terms of overall structural integrity and in terms of robustness against failure of function due to mutations [28].

This kind of network preserves the memory of the protein structure, i.e. it changes if the protein 3D structure changes. In such a way, some topological features, like the structure deformation subsequent to the protein attachment, can be easily described. Furthermore, when the network takes the meaning of some specific physical interaction, also the variation due to a structure change can be described.

In our case, we are interested in giving some measurements characterizing the robustness of the network, so that structure deformation and, subsequently, electric behaviour could be predicted. Those characteristics could have implications on the overall structure, since they can enable a better circulation of knowledge (affinity performance) between the core and the periphery of a network.

In principle, assortativity is directly related to the network robustness in terms of the network connectivity. A failure of a high-degree node (hub), or a targeted attack in a disassortative network would leave other high degree nodes connected to peripherical nodes (resilience) [21]. This minimizes the chance of the network as a whole to become disconnected. In an assortative network, hubs are connected each other. Hence, failure of a high-degree node in an assortative network would have more impact on the connectedness of the network [21].

Networks can evolve through the entry of new nodes that do not connect to any other node (isolates), or through the entry of new nodes that connect to others either by purely random attachment mechanism or by preferential attachment mechanism [18]. Random attachment means that entering nodes randomly connect to others with no particular preference for their position in the structure. Isolate entrants and random attachment mechanism give rise to a rather flat hierarchy of node degrees [16].

Hierarchy-assortativity of the networks can be investigated through two simple statistical measures. The first concerns with the degree distribution of the network. The more sloped the distribution is, the more the network displays hierarchy in the degree of nodes. From highly connected nodes to weakly connected nodes, the degree distribution exemplifies the level of heterogeneity in the network in terms of actual relational capacity. The second corresponds to the degree correlation. Networks can be characterized as assortative or disassortative to the extent that they display a positive or negative correlation degree. A network is assortative when high degree nodes are connected to other high degree nodes, and low degree nodes are preferentially connected to low degree nodes, so that the degree correlation is positive. A network is disassortative when high degree nodes tend to connect to low degree nodes, and *vice versa*, so that the degree correlation is negative [21].

In the present paper, we analyse the evolution of the TBA network as obtained by adding new links and due to *a.* the attachment of thrombin (a network of two networks) and *b.* the larger value of $R_C$. We will observe that while enlarging $R_C$ does not substantially change the network nature, the presence of thrombin produces a sharp change in the assortativity behaviour.

As a first tool of investigation, we use the degree distribution, i.e. the probability P(*k*) of finding a node with *k* (degree) links. The probability function follows a bell-shaped behaviour (Figure 2a) and reveals some specific information about the original structures. First, the native structure is less dilated than the active structures. In fact, there are no nodes with only two links and the degree with maximal probability is six. The largest active structure is that obtained in the presence of potassium ions.

Of course, the degree distribution is a fingerprint of the network in a certain configuration and therefore it changes when the number of links changes. In our model, this is easily obtained by varying the value of the interaction radius $R_C$.

In Figure 2b, the degree distribution is shown for the complex TBA-thrombin in the presence of both sodium and potassium ions, for $R_C$= 11.3 Å. As in Figure 2a, the distribution shows a bell-shaped behaviour, with clear differences between the two structures, in particular, the potassium-structure is found to be the most dilated.

This kind of description could be quantitative in the case of scale-free networks, in which P(k)~ $k^a$ and the value of *a* could be used to identify the universality class of the network. Otherwise, it becomes quite qualitative when the specific shape of the distribution is hard to be exactly identified. Therefore, to obtain a more effective information, we analyse the rank-degree distribution, which is obtained by organizing the node degrees from the largest to the smallest [21]. It has been argued that a close relation does exist between the degree and the rank-degree distribution [29, 30], for scale free network. We have found that when the degree distribution is bell-shaped, the rank-degree distribution is finely fitted by an exponential function, say: $k=A\ exp(ar)$, with $k$ the degree, $r$ the rank of each node, and $A$ a proportionality factor. In particular, in Figure 3 we reproduce the degree distribution of few connected networks, with $R_C$= 11.3 Å. Figure 3a shows the rank-degree distribution for the TBA alone, in the native and active states. We can observe a quite good fit with the exponential behaviour. The values of $a$, within the error of 1% are reported in this figure.

*Corresponding Authors:
eleonora.alfinito@unisalento.it
rosella.cataldo@unisalento.it

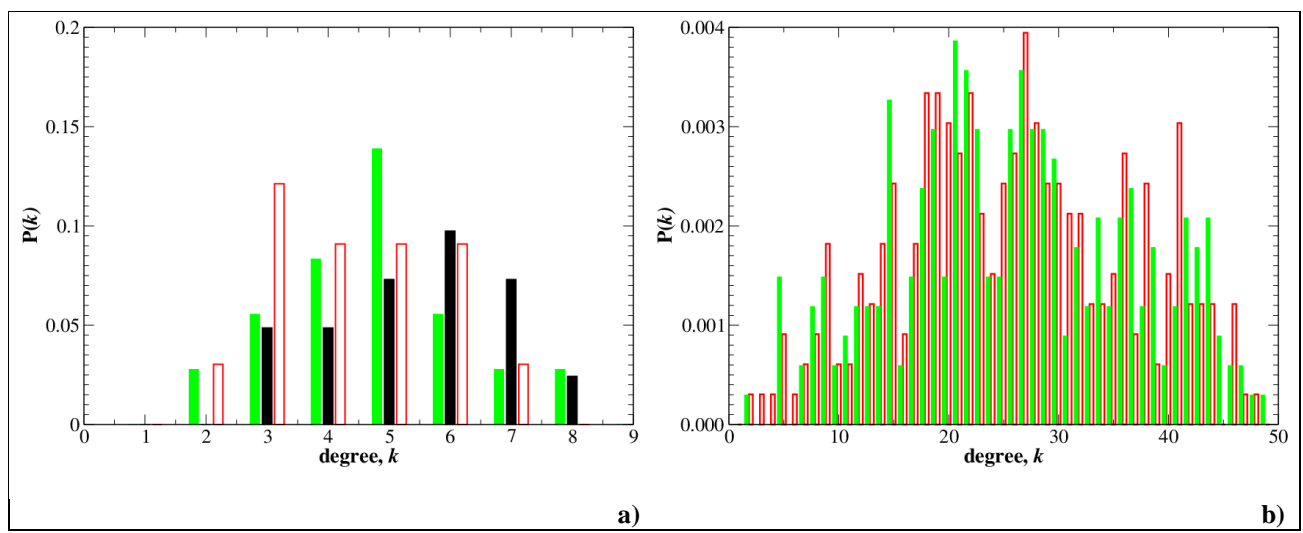

**Figure 2.** *a. Degree distribution of TBA alone in the native state (black full) in the active state with potassium ions, entry 4DII (red empty), and in the active state with sodium ions, entry 4DIH (green full) [15]. b. Degree distribution of TBA complexed with thrombin in the presence of potassium ions, entry 4DII (red empty), and sodium ions, entry 4DIH (green full). The value of $R_C$ is 11.3 Å for both the figures.*

Figure 3b shows the rank-degree distribution for TBA conjugated with thrombin. In particular, it is possible to identify two different behaviours: a former, quite flat, not ion-specific, and covering almost all the nodes; a latter, quite sharp, sensitive to the presence of a specific ion, and for nodes with very low degree. Both behaviours can be described by an exponential law, whose exponent varies with $R_C$. The former region is characterized by: $a$= -0.0047 for $R_C$= 11.3 Å, and $a$=-0.0020 with $R_C$= 30 Å; the latter region is characterized by $a$=-0.077(Na$^+$) and -0.97(K$^+$) for $R_C$= 11.3 Å, and $a$=-0.081(Na$^+$) and -0.087(K$^+$) with $R_C$= 30 Å. All fits have an uncertainty within 2%.

The former region mainly describes the complex core, which is quite uniform and becomes more and more uniform (flat) by enlarging $R_C$. Otherwise, the latter region mainly describes the aptamer and the aptamer-thrombin binding region, which is maximally sensitive to the different ions and is structurally less uniform than the core.

*Corresponding Authors:
eleonora.alfinito@unisalento.it
rosella.cataldo@unisalento.it

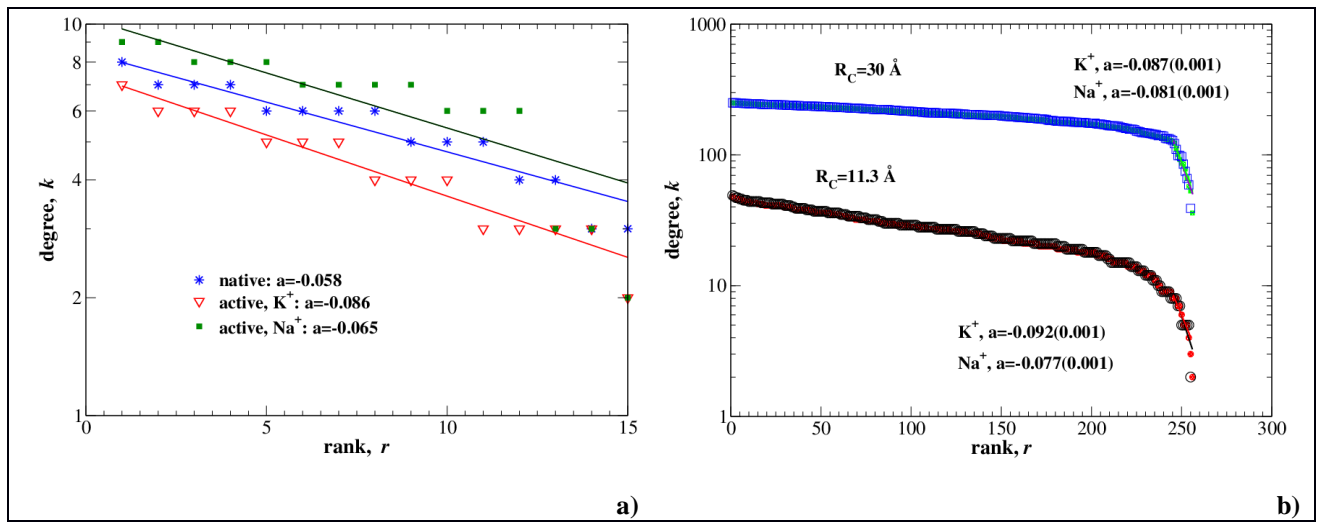

**Figure 3.** *a) Rank-degree distribution of TBA in the native state (stars), in the active state with potassium ions (empty triangles), in the active state with sodium ions (full squares). Lines are the exponential fits. The value of $R_C$ is 11.3 Å. b) Rank-degree distribution of TBA complexed with thrombin, in the presence of potassium ions (empty symbols) and sodium ions (full symbols) for two different values of the interaction radius. Lines are the exponential fits.*

As figures show, the higher $R_C$ the flatter the distribution, since, as expected, this strategy to add links is uniform and hides the differences between nodes (the rank). As a matter of fact, usually $R_C$ value is sufficiently large to have a connected network but not so large to hide the node peculiarities [31].

Figure 4 show the degree correlation [21] for the aptamer conjugated with thrombin, and (in the insets) in the active state, in the presence of potassium *ions* (Figure 4a) and sodium *ions* (Figure 4b). Degree correlation states a relation between the mean degree *[k]* of the nearest neighbors of each node and the degree *k* of that node [21]. The neighbors of the *l*-th node correspond to the non-zero elements of the *l*-th row of the adjacency matrix. The linear behaviour *[k]=D+bk*, with *D* a fitting constant, well fits the data. We can observe that the value of *b* is strictly positive (assortativity) only for the complex, while the aptamer alone has a null or negative slope (disassortativity). This is a quite intriguing result since it sheds light on the way in which the protein connects to the aptamer, in other terms, to the affinity. In principle, a disassortative network is a network able to exchange information with the environment, what we can call an open system, otherwise, an assortative network is a closed system, because information is trapped in regions of similar connection levels. It is stratified. Therefore, adding the protein the system becomes closed, more stable and less affected by external stimuli. This could be identified within the network approach, as an effective definition of affinity.

The slope value depends on the value of $R_C$. By enlarging $R_C$, nodes with a larger degree connect with nodes with a lower degree, so that although the aptamer alone and the aptamer complexed have different assortativity, new links are anyway added in order to reduce the assortativity level. Similar results have been obtained in the presence of sodium ions, as shown in Figure 4b. We observe that the values of the slope are not exactly the same shown in Figure 4a, mainly for the aptamer alone, which suggests that the affinity performances could be monitored by looking at this quantity.

In conclusion, the topological analysis reveals that the differences between the macromolecules in the presence of the cations $K^+$ and $Na^+$ are quite subtle, as expected from previous investigations [14, 15]. Anyway, the hierarchy and assortativity analysis give some insights on the way to quantify these differences.

A very effective way to explore the aptamer features is also given by the electrical response, as will be reported in the following section.

*Corresponding Authors:
eleonora.alfinito@unisalento.it
rosella.cataldo@unisalento.it

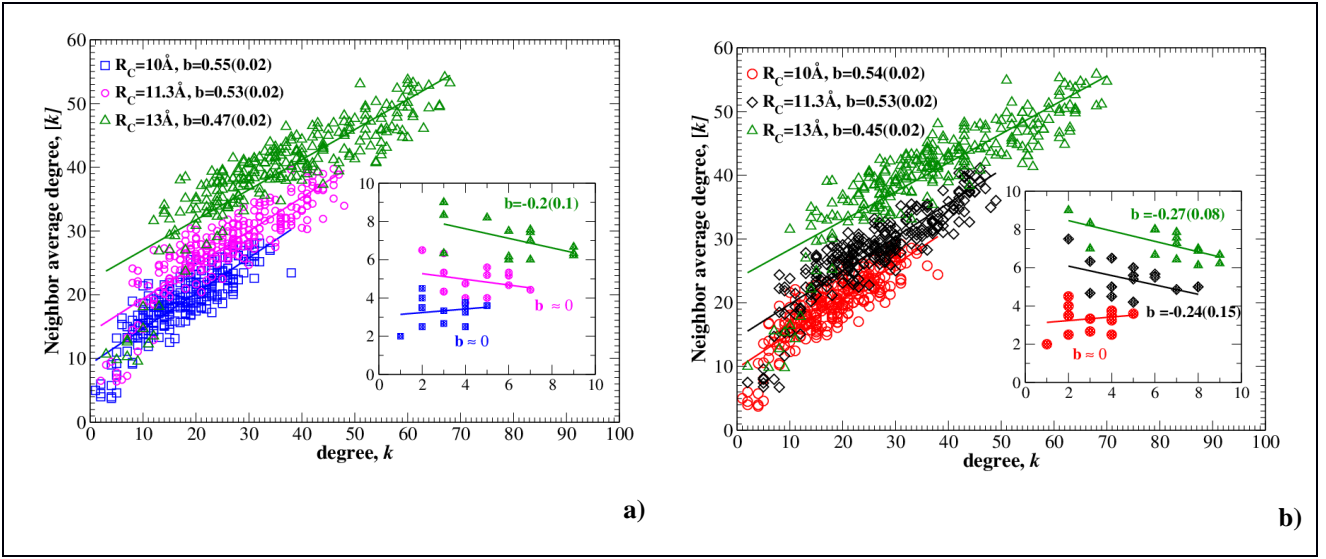

**Figure 4.** *a) Degree correlation of TBA complexed with thrombin or in the active state ( in the inset) in the presence of potassium ions, for three different value of the interaction radius. Lines are the linear fits, the standard error is given in brackets. b) Degree correlation of TBA complexed with thrombin, or in the active state (in the inset) in the presence of sodium ions, for three different value of the interaction radius. Lines are the linear fits, the standard error is given in brackets.*

### 2.2 Electrical description

Each network can be tailored with the electrical features specific of the corresponding macromolecules.
In doing so, the physico-chemical interactions we are interested to describe, are introduced by attributing to each link the corresponding properties at the elemental level. In the present case, an electromagnetic interaction takes into account the response of the protein to different electrical solicitations.

An exhaustive argumentation of this item is given in Alfinito et al. [14], in which a resistance measurement has been proposed as an efficient tool for testing different affinities in aptamers. Here the main principles are considered and compared with the findings reported in the previous section and in the literature insert reference.
In particular, we aim to describe the results of some EIS measurements [13], which reveal a change in resistance and capacitance of a sample of TBA when incubated with different concentrations of thrombin. Therefore, we dress our network with the appropriate electrical elementary impedances, i.e. those able to reproduce, as a global macromolecule response, that observed in experiments. In particular, it writes:

$$Z_{i,j} = \frac{S_{i,j}\rho_{i,j}}{1+i\rho_{i,j}\ \varepsilon_{i,j}\varepsilon_0\ \omega} \tag{1}$$

To formulate this expression, we notice that the observed impedance variation is well reproduced by an RC parallel circuit [3, 31], here $\rho_{i,j}$ and $\varepsilon_{i,j}$ represent the resistivity and the dielectric constants of the link connecting two neighbor bricks [14]. The shape size of the equivalent element is resumed by $S_{i,j}$, with dimension inverse of a length and which depends on the distance between the bricks and the interaction radius. Finally, ω is the circular frequency of the applied bias (voltage or current). The network is connected to the external bias by means of ideal contacts located on the first and last node, and is electrically solved by adopting standard techniques, as detailed in the following.

The node Kirchhoff law allows to write down a set of linear equation for the $v_i$ variables (the electrical potential of node *i*). The total network impedance is calculated solving the electrical network within the standard Gauss–Jordan method [32] and using the constraints $v_1=0$ and $v_N=U$, with *U* the bias value. The process is repeated for different frequencies and/or $R_C$ values. In such a way, we are able to produce two kinds of spectra, the frequency and the interaction spectra [3, 11, 31].

*Corresponding Authors:
eleonora.alfinito@unisalento.it
rosella.cataldo@unisalento.it

We distinguish among the resistance of TBA, by using the same notation of Sec.2.1:
- in the native state (a), hereafter called $r_{nat}$ ;
- in the active state (b), hereafter called $r_{act}$ ;
- complexed (c), hereafter called $r_{com}$.

This procedure allows for the calculation of the capacitance and resistance variation of a single as well as of a sample of macromolecules [14, 33, 34]. As expected, the results depend on the value of $R_C$, which, in turn, is related to the value of the macromolecule free energy [27, 34, 35]. Therefore, a complete information is obtained by analyzing the interaction spectra, i.e. the set of data obtained over a continuous range of $R_C$ values.

In particular, we calculate the interaction spectrum of the resistance mean value over the 12 realizations of the aptamer in the native state, as given by the Protein Data Bank entry 148D [25, 26]. As shown in Figure 5, there is a very large increase of the resistance of the aptamer in the active state with respect to the aptamer in the native state.

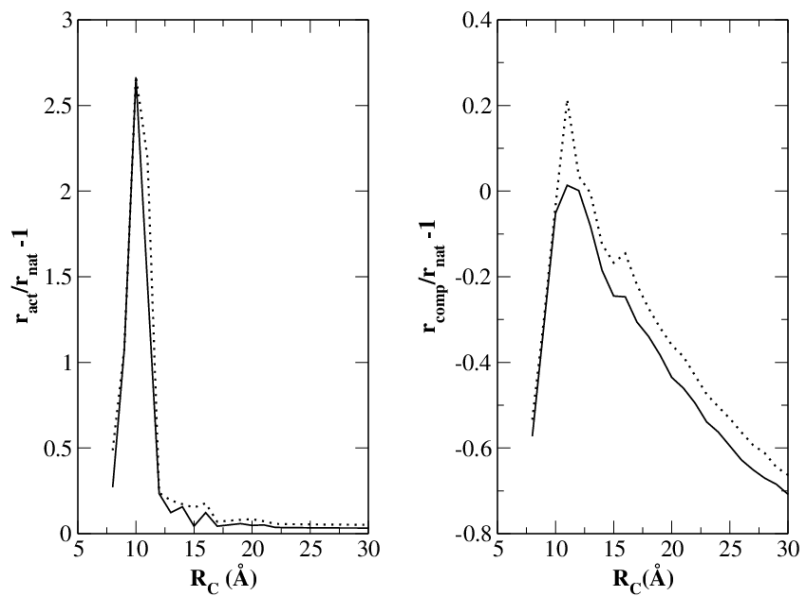

**Figure 5.** *Relative resistance spectrum of the aptamer in the active state (figure on the left) and complexed with thrombin (figure on the right). The resistance is compared to that of the native state. Dotted (continuous) curves refer to the aptamer in the presence of potassium (sodium) ions, respectively.*

This result does not depend significantly on the kind of ions in solution and is in agreement with the more dilated structure of the active state, observed in the topological analysis. Notice that the sensitive region is of a few Angstroms, centred around 10 Å. By adding the protein, the difference between the structure in sodium or potassium ions becomes quite relevant. In particular, the response of the complex in sodium ions exhibits a negligible difference with respect to the response expected for the native state. On the other side, a resistance increase up to 20% (at 11.3 Å) is predicted in the presence of potassium ions [14].

As a final remark, the presence of thrombin strongly reduces the complex resistance. This is a consequence of the very large structure of the aptamer and the very good way in which the thrombin docks it. The anchoring is very efficient and produces many links which reduce the global resistance. In the case of the complex in sodium ions, the resistance of the aptamer reduces to that of the native state, in the case of the complex in potassium ions, the complex has a higher difference with respect to the native state. In other terms, and as shown in the analysis of the topological aspects, this structure is a little bit larger than the other. A resistance increase after the conjugation with thrombin has been indeed observed in experiments [13].

## 3. Discussion

Now we will attempt to discuss the results of the previous section, with the aim to give a more precise characterization of the affinity performance of the network, representing the aptamer-protein complex.

*Corresponding Authors:
eleonora.alfinito@unisalento.it
rosella.cataldo@unisalento.it

To the best of our knowledge, it is the first time that a network approach is applied to the aptamer-protein complex, and found to reveal peculiar characteristics. In the literature, in fact, many works limited the analysis to protein (protein-protein) networks, highlighting different behaviour in network topology and statistical measures of the structures.

In del Sol et al. [36] protein structures (monomers and dimers) were modelled as small-world network, in accordance with their values of clustering coefficients and characteristic path lengths, in comparison with random and regular graphs with the same number of vertices and average number of neighbours. The frequency of the residue number of links $N$, averaged in both sets of monomers and dimers, followed a Poisson-like distribution, and by using only one network topology characteristic (betweenness centrality) were able to identify hot spot regions, at protein–protein interfaces.

Tanaka et al. [29] analysed protein–protein scale-free networks, in which the rank-degree function, $r = f(d)$, and the degree distribution $P(k)$ followed a power law degree distribution. Indeed, in this regard Wu et al. [30] showed that those networks are scale-free, when the scaling exponent in the power law rank-degree function is greater than 2, concluding that a mathematical theoretical framework on scale-free networks is expected [30]. However, by using both frequency–degree and rank–degree plot, Wu et al. [30] argued important features, confirming such a statistic as a reliable analytic tool for protein-protein networks.

Taylor [28] proposed a review of many papers that analysed protein structures by using small world, with the aim of building a quantitative model for predicting ligand binding affinity. Characteristics, such as closeness, path length and clustering coefficient were considered to maximise potential small-world network and, incorporated into existing models, providing better quality predictions of *structure and function*.

In our case (aptamer-protein), network topology is compatible with the hypothesis of a small-world network [18], with significant clusterization. The degree distribution $P(k)$ follows a bell-shaped behaviour. The rank-degree distribution for the TBA alone, in the native and active states, and complexed with thrombin exhibits an exponential behaviour (Figure 3).

Figure 6 resumes on a single plot the results of the research, as regards hierarchy-assortativity. Specifically, each point of coordinates ($b$, $-a$) describes the aptamer in a specific state, at $R_C$=11.3 Å. This kind of plot gives an instant sketch of the differences between the aptamer alone, in the left semi-plane and the aptamer complexed, in the right semi-plane.

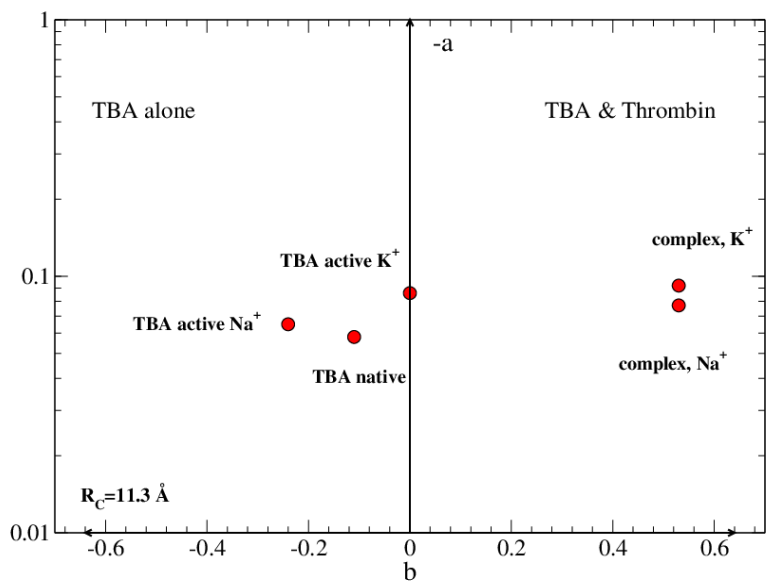

**Figure 6.** *Hierarchy-assortativity plot of TBA in the native, active and thrombin complexed states ($R_C$ value is 11.3 Å).*

TBA, in the native state, or in presence of K$^+$ and Na$^+$ ions, presents a quite low hierarchical organization ($-a \leq 0.1$) and a degree correlation $b$, strictly non-positive (Figure 6). This classifies the networks as quite flat and disassortative. In other terms, they are "open" networks, since core and periphery nodes are connected between them, thus allowing a good circulation of information from inside to outside. From this point of view ($b < 0$), sodium allows a better information flow with respect potassium ($b \approx 0$), and produces a more robust (flat) network (smaller $a$). On the other side, this also means that the network does not need to be completed to be stable (see the native state) i.e. this network is less inclined to accept an external target (low affinity).

*Corresponding Authors:
eleonora.alfinito@unisalento.it
rosella.cataldo@unisalento.it

Looking at the complex structures, they have the same, positive, value of *b*, i.e. they both show an assortativity behaviour, i.e. information is stratified and this does not allow a good circulation, the system is closed, nothing more can be added to the structure. Furthermore, they have a quite similar value of *a*, which still classifies them as flat networks, in other terms, they are quite robust against random attacks or failures.

In conclusion, by adding the thrombin, the networks go from open to closed (in terms of structure and also of affinity performance) systems and this is the most impressive news given by Figure 6.

To complete our investigation we have tried to understand whether the assortativity transition observed in TBA is due to its peculiar G-quadruplex structure or is a more general feature. Although at present there is a few information about the 3D structure of aptamers, we have the possibility to calculate assortativity for a set of anti-angiopoietin macromolecules, whose structure has been computationally obtained. The result, shown in Figure 7, is that these aptamers show substantially the same assortativity, which is different from that of TBA.

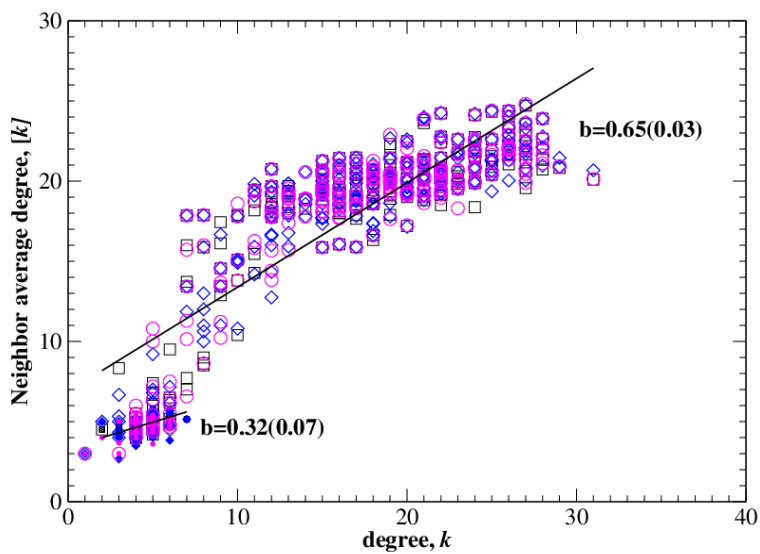

**Figure 7.** *Degree correlation of a set of anti-angiopoietin aptamers [6, 37].*

## 4. Conclusions

We have developed a model for analysing affinity performance of the complex constituted by the aptamer TBA and its specific ligand, the thrombin enzyme. The inhibition activity of TBA on this protein is a long time known result, actually extensively investigated to produce a targeted therapy with reduced side effects and also considered for thrombin biosensors production. From one hand, to the best of our knowledge, it is the first time that network science is applied to the aptamer-protein complex, revealing peculiar characteristics. From another hand, the literature proposes many works analysing protein (protein-protein) networks, highlighting different behaviours in network topology and describing some statistical features of the structures.

The developed model encompasses two fundamental steps: the graph analogue building and the interaction network building. Both steps give useful data for the system investigation. The topological analysis reveals different affinity performances between the macromolecules in the presence of two different cations, as expected by previous investigations in literature [14, 15]. In particular, the hierarchy-assortativity measurements give quite intriguing results in terms of the affinity performance of the complex.

By deriving the electrical network from the topological one, we have observed that the electrical resistances vary when TBA is complexed with $Na^+$ or $K^+$, confirming the relevant role of the cations in the binding mechanism, according to the literature.

*Corresponding Authors:
eleonora.alfinito@unisalento.it
rosella.cataldo@unisalento.it

Independently of the specific results, the emphasis is placed on the principles of the model that can be generalized to evaluate binding affinity in other aptamer-ligand complexes. In a more pragmatic approach, these results suggest that a measure of resistance could be an indicator of affinity performance.

From the above, the graph topological analysis emerges as a novel theoretical tool able to validate hierarchy and assortativity as relevant quantities for investigating affinity.

*Corresponding Authors:
eleonora.alfinito@unisalento.it
rosella.cataldo@unisalento.it

*Corresponding Authors:
eleonora.alfinito@unisalento.it
rosella.cataldo@unisalento.it